\documentclass[12pt,reqno]{amsart}

\usepackage{amsmath,amsfonts,amsthm,amssymb}

\usepackage[a4paper,height=20cm,top=24mm,bottom=24mm,left=26mm,right=26mm]{geometry}

\usepackage{hyperref}

\usepackage[]{graphicx}                                                    
\usepackage{booktabs}
\usepackage[table]{xcolor}

 \usepackage{titlesec}
 \titleformat{\section}
 {\bfseries\scshape\centering}
 {\thesection.}{.5em}{}
 \titleformat{\subsection}
 {\rmfamily\bfseries}
 {\thesubsection}{.5em}{}
 \titleformat{\subsubsection}
 {\rmfamily\bfseries}
 {\thesubsubsection}{.5em}{}
\numberwithin{equation}{section}

\newcommand{\I}{\mathrm{i}}
\newcommand{\E}{\mathrm{e}}

\DeclareMathOperator{\Tr}{Tr}

\DeclareMathOperator{\ch}{ch}

\DeclareMathDelimiter{\Norm}{\mathord}{largesymbols}{"3E}{largesymbols}{"3E}

\DeclareMathOperator{\ord}{ord}
\DeclareMathOperator{\mult}{mult}
\DeclareMathOperator{\Mat}{Mat}
\DeclareMathOperator{\Cusp}{Cusp}

\allowdisplaybreaks[1]
\begin{document}
\baselineskip 16pt
\parskip 8pt
\sloppy


\title{Twisted Elliptic Genus for $K3$ and Borcherds Product}


\author[T. Eguchi]{Tohru \textsc{Eguchi}}

\author[K. Hikami]{Kazuhiro \textsc{Hikami}}


\address{Yukawa Institute for Theoretical Physics,
  Kyoto University,
  Kyoto 606--8502, Japan,
  \indent California Institute of technology,
  E. 1200 California Bld.,
  Pasadena CA 91125,
  \indent U.S.A. }

\email{
  \texttt{eguchi@yukawa.kyoto-u.ac.jp}
}

\address{Faculty of Mathematics,
  Kyushu University,
  Fukuoka 819-0395, Japan.}

\email{
  \texttt{KHikami@gmail.com}
}


\date{December 26, 2011.  Revised  on May 7, 2012.}

\begin{abstract}
We further discuss the relation between the elliptic genus of 
$K3$ surface and the Mathieu group $M_{24}$.
We find that some of the twisted elliptic genera for $K3$ surface,
defined for conjugacy classes of the Mathieu group
$M_{24}$, can be represented in a very simple manner in terms of the 
$\eta$ product of the corresponding conjugacy classes.
It is shown that 
our formula is a consequence of the identity between the Borcherds product and 
additive lift of some Siegel modular forms.

\end{abstract}





\maketitle


\section{Introduction}

Studies of the $K3$ surface based on the representation theory of 
superconformal algebras (SCA) was initiated some time ago \cite{EguOogTaoYan89a}.
Therein, the elliptic genus of $K3$ surface was decomposed into a
sum of characters of $\mathcal{N}=4$ SCA. 
Recently, an intriguing property of this decomposition was discovered 
~\cite{EgucOoguTach10a};
multiplicities of the non-BPS 
representations are given by the sum of dimensions of 
irreducible representations of the 
Mathieu group $M_{24}$.
This observation is a variant of the famous Monstrous
moonshine~\cite{ConwayNorton79} where the Fourier coefficients of the 
modular $J$-function is expressed as the sum of dimensions of
representations of the Monster group. 
In our case of  ``Mathieu moonshine'',   
multiplicities of non-BPS representations are Fourier 
coefficients of a certain mock theta function~\cite{EguchiHikami08a,EguchiHikami09a}.

The Mathieu group $M_{24}$ has 26 conjugacy classes. 
To each conjugacy class we can introduce a twisted version of the 
elliptic genus of $K3$ surface which is an analogue of McKay--Thompson
series in 
the Monstrous moonshine. 
One can determine uniquely the decomposition of the multiplicities of
non-BPS representations  once one has a complete list of twisted
genera for all conjugacy  classes. 

Recently a completed list of twisted genera became
available~\cite{MCheng10a,GabeHoheVolp10a},~\cite{GabeHoheVolp10b,EguchiHikami10b}
and by 
making use them the decomposition has been carried out up to a very high 
level $\approx$1000. 
This result produces a very strong support for the Mathieu moonshine conjecture.

In the study of the Mathieu group $M_{24}$,  
the $\eta$-products associated with various cycle shapes
play an important role ~\cite{DummKisiMcKa85a,Koike84a,Mason85a}.
Purpose of this paper is to present simple relationships between the twisted
elliptic genera of $K3$ and the $\eta$-products for various conjugacy
classes of $M_{24}$. 
Our result follows from the identity of Siegel modular forms of degree-$2$ 
which are expressed both as an infinite sum (additive lift)
and as an infinite 
product (Borcherds lift) simultaneously.

The Siegel modular form is physically interpreted as the partition function of 
${1\over 4}$ BPS states in the CHL
model~\cite{ChauHockLykk95a,DijkVerlVerl97b} of a heterotic string
theory compactified on $K3\times S^2$.
The twisted elliptic genera of $K3$ for conjugacy classes
$\mathrm{2A}$, $\mathrm{3A}$, $\mathrm{5A}$, $\mathrm{7A}$ are closely 
related to 
the  $\mathbb{Z}_N$ orbifold partition function
of CHL models~\cite{DabhoNampu07a,GovinKrish09b,JatkarSen05,DavidJatkarSen06}.

This paper is organized as follows:
in section~\ref{sec:twist_genus}
we recall the twisted elliptic genera.
By use of the Hecke operator, we express the twisted elliptic genus
in terms of the $\eta$-product for a number of conjugacy classes.
We show that this identity can be derived by using the Borcherds product in
section~\ref{sec:Borcherds}.
The last section is devoted to concluding remarks.
Notations on  modular forms are summarized in the Appendix.

\begin{table}
  \newcolumntype{L}{>{$}l<{$}}
  \newcolumntype{C}{>{$}c<{$}}
  \rowcolors{2}{gray!11}{}
  \centering
  \resizebox{.97\textwidth}{!}{
  \begin{tabular}{cLL}
    \toprule
    $g$ & \text{cycle shape} & \text{permutation}
    \\
    \midrule \midrule
    1A & 1^{24} & ()
    \\
    2A &1^8 \cdot 2^8 & 
    (1,8)(2,12)(4,15)(5,7)(9,22)(11,18)(14,19)(23,24)
    \\
    3A & 1^6 \cdot 3^6 &
    (3,18,20)(4,22,24)(5,19,17)(6,11,8)(7,15,10)(9,12,14)
    \\
    5A & 1^4 \cdot 5^4&
    (2,21,13,16,23)(3,5,15,22,14)(4,12,20,17,7)(9,18,19,10,24)
    \\
    4B &1^4 \cdot 2^2 \cdot 4^4 &
    (1,17,21,9)(2,13,24,15)(3,23)(4,14,5,8)(6,16)(12,18,20,22)
    \\
    7A &1^3 \cdot 7^3 &
    (1,17,5,21,24,10,6)(2,12,13,9,4,23,20)(3,8,22,7,18,14,19)
    \\
    7B&1^3 \cdot 7^3 &
    (1,21,6,5,10,17,24)(2,9,20,13,23,12,4)(3,7,19,22,14,8,18)
    \\
    8A &1^2 \cdot 2^1 \cdot 4^1 \cdot 8^2 &
    (1,13,17,24,21,15,9,2)(3,16,23,6)(4,22,14,12,5,18,8,20)(7,11)
    \\ 
    6A &1^2 \cdot 2^2 \cdot 3^2 \cdot 6^2 &
    (1,8)(2,24,11,12,23,18)(3,20,10)(4,15)(5,19,9,7,14,22)(6,16,13)
    \\
    11A&1^2 \cdot 11^2 &
    (1,3,10,4,14,15,5,24,13,17,18)(2,21,23,9,20,19,6,12,16,11,22)
    \\
    15A &1^1 \cdot 3^1 \cdot 5^1 \cdot 15^1 &
    (2,13,23,21,16)(3,7,9,5,4,18,15,12,19,22,20,10,14,17,24)(6,8,11)
    \\
    15B& 1^1 \cdot 3^1 \cdot 5^1 \cdot 15^1 &
    (2,23,16,13,21)(3,12,24,15,17,18,14,4,10,5,20,9,22,7,19)(6,8,11)
    \\
    14A &1^1 \cdot 2^1 \cdot 7^1 \cdot 14^1 &
    (1,12,17,13,5,9,21,4,24,23,10,20,6,2)(3,18,8,14,22,19,7)(11,15)
    \\
    14B& 1^1 \cdot 2^1 \cdot 7^1 \cdot 14^1 &
    (1,13,21,23,6,12,5,4,10,2,17,9,24,20)(3,14,7,8,19,18,22)(11,15)
    \\
    23A &1^1 \cdot 23^1 &
    (1,7,6,24,14,4,16,12,20,9,11,5,15,10,19,18,23,17,3,2,8,22,21)
    \\
    23B& 1^1 \cdot 23^1 &
    (1,4,11,18,8,6,12,15,17,21,14,9,19,2,7,16,5,23,22,24,20,10,3)
    \\
    \midrule
    12B &12^2 &
    (1,12,24,23,10,8,18,6,3,21,2,7)(4,9,11,15,13,16,20,5,22,17,14,19)
    \\
    6B &6^4 &
    (1,24,10,18,3,2)(4,11,13,20,22,14)(5,17,19,9,15,16)(6,21,7,12,23,8)
    \\
    4C &4^6 &
    (1,23,18,21)(2,12,10,6)(3,7,24,8)(4,15,20,17)(5,14,9,13)(11,16,22,19)
    \\
    3B &3^8 &
    (1,10,3)(2,24,18)(4,13,22)(5,19,15)(6,7,23)(8,21,12)(9,16,17)(11,20,14)
    \\
    2B &2^{12} &
    (1,8)(2,10)(3,20)(4,22)(5,17)(6,11)(7,15)(9,13)(12,14)(16,18)(19,23)(21,24)
    \\
    10A &2^2 \cdot 10^2 &
    (1,8)(2,18,21,19,13,10,16,24,23,9)(3,4,5,12,15,20,22,17,14,7)(6,11)
    \\
    21A &3^1 \cdot 21^1 &
    (1,3,9,15,5,12,2,13,20,23,17,4,14,10,21,22,19,6,7,11,16)(8,18,24)
    \\
    21B& 3^1 \cdot 21^1 &
    (1,12,17,22,16,5,23,21,11,15,20,10,7,9,13,14,6,3,2,4,19)(8,24,18)
    \\
    4A &2^4 \cdot 4^4 &
    (1,4,8,15)(2,9,12,22)(3,6)(5,24,7,23)(10,13)(11,14,18,19)(16,20)(17,21)
    \\
    12A &2^1 \cdot 4^1 \cdot 6^1 \cdot 12^1 &
    (1,15,8,4)(2,19,24,9,11,7,12,14,23,22,18,5)(3,13,20,6,10,16)(17,21)
    \\
    \bottomrule
  \end{tabular}
}
  \caption{Representatives of conjugacy classes $g$.}
  \label{tab:class}
\end{table}

\section{Twisted Elliptic Genus and Cycle Shape}
\label{sec:twist_genus}
\subsection{Character Decomposition}

The elliptic genus of $K3$ surface is known to be a Jacobi form
with weight $0$ and index $1$, and it is decomposed as
\begin{equation}
  \label{decompose_Z_K3}
  Z_{K3}(z;\tau)
  =
  20 \ch^{\widetilde{R}}_{\frac{1}{4},0}(z;\tau)
  - 2 \, \ch^{\widetilde{R}}_{\frac{1}{4}, \frac{1}{2}}(z;\tau)
  +
  \sum_{n=1}^\infty A(n) \, 
  \ch^{\widetilde{R}}_{n+\frac{1}{4}, \frac{1}{2}}(z;\tau) ,
\end{equation}
where $\ch^{\widetilde{R}}_{{1\over 4},\ell}(z;\tau)$,
($\ell=0$, $1/2$),
$\ch^{\widetilde{R}}_{{n+{1\over 4}},{1\over 2}}(z;\tau)$ are characters
of BPS and non-BPS representations of $\mathcal{N}=4$ SCA in $R$
sector (with $(-1)^F$ insertion).
$\ell$ denotes the iso-spin.
See Appendix for their explicit forms.
$A(n)$ is the multiplicity of the non-BPS representation with
$h=n+1/4$,
and is given as follows;
\begin{equation*}
  \begin{tabular}{rrrrrrrrrrr}
    \toprule
    $n$ & 1 & 2 & 3 & 4 & 5 & 6 & 7 & 8 & 9 & $\cdots$
    \\
    \midrule
    $A(n)$ & 90 & 462 & 1540 & 4554 & 11592 & 27830 & 61686 & 131100 &
    265650 & $\cdots$
    \\
    \bottomrule
  \end{tabular}
\end{equation*}

Numbers $A(n)$ are 
the expansion coefficients of a certain mock theta function with a
shadow $\eta(\tau)^3$~\cite{EguchiHikami08a}.
Observation in~\cite{EgucOoguTach10a} is that 
$A(n)$ is decomposed
into a sum of dimensions of
the irreducible representations $R$ of $M_{24}$,
\begin{equation}
  \sum_R \mult_R(n) \, \dim R = A(n) ,
\end{equation}
and that the multiplicities $\mult_R(n)$ are conjectured to be positive integers.
By introducing a vector space
\begin{equation}
  V(n) = \bigoplus_R \mult_R(n) \, R ,
\end{equation}
we can write $A(n)$ as 
\begin{equation}
  A(n) = \Tr_{V(n)} 1 .
\end{equation}
\subsection{Twisted Elliptic Genus of $K3$}

Corresponding to each conjugacy class $g$ of the  Mathieu group $M_{24}$ 
we can define a variant of   $A(n)$ by
\begin{equation}
  \begin{aligned}[b]
    A_g(n) &= \Tr_{V(n)} g
    = \sum_R \mult_R(n) \, \chi_R^{\,\,\,g} .
  \end{aligned}
\end{equation}
Here $\chi_R^{\,\,\,g}=\Tr_R \,g$ denotes the character of $M_{24}$
for the representation $R$ and conjugacy class $g$. 
We define the twisted elliptic genus 
$Z_g(z;\tau)$ for a class $g$ 
by the decomposition
\begin{align}
  Z_g(z;\tau)
  & =
  \left( \chi_g - 4 \right) \,
  \ch^{\widetilde{R}}_{h=\frac{1}{4},\ell=0}(z;\tau)
  - 2 \,
  \ch^{\widetilde{R}}_{h=\frac{1}{4},\ell=\frac{1}{2}}(z;\tau)
  +
  \sum_{n=1}^\infty
  A_g(n) \,
  \ch^{\widetilde{R}}_{h=n+\frac{1}{4},\ell=\frac{1}{2}}(z;\tau) 
  \nonumber
  \\
  &=
  \frac{
    \left[ \theta_{11}(z;\tau) \right]^2}{
    \left[\eta(\tau) \right]^3} 
  \left(
    \chi_g \,
    \mu(z;\tau)
    -
    \Sigma_g(\tau) 
    \right) .
  \label{twisted_decompose}
\end{align}
Here
$\chi_g\in \mathbb{Z}$ is the Witten index of twisted genus,
\emph{i.e.}
$Z_g(z=0;\tau)$.

We have used
\begin{equation}
  -q^{\frac{1}{8}} \, \Sigma_g(\tau)
  =
  -2 + \sum_{n=1}^\infty A_g(n) \, q^n .
\end{equation}
This decomposition reduces to \eqref{decompose_Z_K3} 
when $g=\mathrm{1A}$.

We classify conjugacy classes into type~I and type~II;
conjugacy classes of type~I contain a cycle of $\text{length}=1$,  
\emph{i.e.} 
a fixed point under permutation of 24 elements and thus they 
arise from those of $M_{23}$.
See Table~\ref{tab:class}.
On the other hand conjugacy classes of type~II are those which are
intrinsically $M_{24}$. These two types of conjugacy classes have a
qualitatively different behavior. 
It should be noted that, in our following studies on twisted elliptic
genera,
there is no difference between
conjugacy classes with the same cycle shape such as 
$\mathrm{7A}$ and $\mathrm{7B}$,
and we
use  a notation $\mathrm{7AB}$ for such a pair of conjugacy classes.

The twisted elliptic genera were first obtained for (mainly) type~I
classes~\cite{MCheng10a,GabeHoheVolp10a} and then obtained for type~II
classes~\cite{GabeHoheVolp10b,EguchiHikami10b},
and their complete
list is
available now.
See Table~\ref{tab:twisted_genus}.

It turns out that their Witten indices are given by
\begin{equation*}
  \newcolumntype{C}{>{$}c<{$}}
  \centering
  \begin{tabular}{C||*{12}{C}|C}
    \toprule
    & \multicolumn{12}{C|}{
      \text{type I}
    }
    &
    \text{type II}
    \\    
    g &
    \mathrm{1A} & \mathrm{2A} & \mathrm{3A} & \mathrm{5A} &
    \mathrm{4B} &  \mathrm{7AB} & \mathrm{8A} & \mathrm{6A} &
    \mathrm{11A} & \mathrm{15AB} & \mathrm{14AB} & \mathrm{23AB} &
    \text{others}
    \\
    \midrule
    \chi_g &
    24 & 8 & 6 & 4 & 4 & 3 & 2 & 2 & 2 & 1 & 1 & 1 & 0
    \\
    \bottomrule
  \end{tabular}
\end{equation*}
Note that 
\begin{equation}
\chi_g=\Tr_{\rho_1}g+\Tr_{\rho_{23}}g
\end{equation}
where $\rho_1,\rho_{23}$ denote 1 and 23-dimensional representation of
$M_{24}$, respectively.

Hereafter we mainly work with  
the conjugacy classes $g$ of type~I
which possess non-vanishing indices 
$\chi_g \neq 0$.

\begin{table}
  \newcolumntype{L}{>{$\displaystyle }l<{$}}
  \rowcolors{2}{gray!11}{}
  \centering
  \resizebox{0.97\textwidth}{!}{
    \begin{tabular}{cL}
      \toprule
      $g$ & Z_g(z;\tau)
      \\
      \midrule \midrule
      1A & 2 \, \phi_{0,1}(z;\tau)
      \\[2mm]
      2A &
      \frac{2}{3} \, \phi_{0,1}(z;\tau)
      + \frac{4}{3} \, \phi_2^{(2)}(\tau) \,
      \phi_{-2,1}(z;\tau)
      \\[2mm]
      3A &
      \frac{1}{2} \, \phi_{0,1}(z;\tau)
      + \frac{3}{2} \, \phi_{2}^{(3)}(\tau) \, \phi_{-2,1}(z;\tau)
      \\[2mm]
      5A &
      \frac{1}{3} \, \phi_{0,1}(z;\tau)
      +
      \frac{5}{3} \, \phi_2^{(5)}(\tau) \, \phi_{-2,1}(z;\tau)
      \\[2mm]
      7AB &
      \frac{1}{4} \, \phi_{0,1}(z;\tau)
      +
      \frac{7}{4} \, \phi_2^{(7)}(\tau) \, \phi_{-2,1}(z;\tau)
      \\[2mm]
      4B &
      \frac{1}{3} \, \phi_{0,1}(z;\tau)
      + \left(
        - \frac{1}{3} \, \phi_2^{(2)}(\tau)
        + 2 \, \phi_2^{(4)}(\tau)
      \right) \, \phi_{-2,1}(z;\tau)
      \\[2mm]
      6A &
      \frac{1}{6} \, \phi_{0,1}(z;\tau)
      + \left(
        - \frac{1}{6} \, \phi_2^{(2)}(\tau)
        - \frac{1}{2} \, \phi_2^{(3)}(\tau)
        + \frac{5}{2} \, \phi_2^{(6)}(\tau)
      \right) \, \phi_{-2,1}(z;\tau)
      \\[2mm]
      8A &
      \frac{1}{6} \, \phi_{0,1}(z;\tau)
      + \left(
        - \frac{1}{2} \, \phi_2^{(4)}(\tau)
        + \frac{7}{3} \, \phi_2^{(8)}(\tau)
      \right) \, \phi_{-2,1}(z;\tau)
      \\[2mm]
      11A &
      \frac{1}{6} \, \phi_{0,1}(z;\tau)
      + \left(
        \frac{11}{6} \, \phi_2^{(11)}(\tau)
        - \frac{22}{5} \, 
        \left[ \eta(\tau) \, \eta(11 \, \tau) \right]^2
      \right) \, \phi_{-2,1}(z;\tau)
      \\[2mm]
      14AB &
      \frac{1}{12} \, \phi_{0,1}(z;\tau)
      + \left(
        -\frac{1}{36} \, \phi_2^{(2)}(\tau)
        -\frac{7}{12} \, \phi_2^{(7)}(\tau)
        +\frac{91}{36} \, \phi_2^{(14)}(\tau)
        - \frac{14}{3} \, 
        \eta(\tau) \, \eta(2 \, \tau)  \, \eta(7 \,\tau)
        \, \eta(14 \, \tau)
      \right) \, \phi_{-2,1}(z;\tau)
      \\[2mm]
      15AB &
      \frac{1}{12} \, \phi_{0,1}(z;\tau)
      + \left(
        -\frac{1}{16} \, \phi_2^{(3)}(\tau)
        -\frac{5}{24} \, \phi_2^{(5)}(\tau)
        +\frac{35}{16} \, \phi_2^{(15)}(\tau)
        - \frac{15}{4} \, 
        \eta(\tau) \, \eta(3 \, \tau)  \, \eta(5 \,\tau)
        \, \eta(15 \, \tau)
      \right) \, \phi_{-2,1}(z;\tau)
      \\[2mm]
      23AB &
      \frac{1}{12} \, \phi_{0,1}(z;\tau)
      +
      \left(
        \frac{23}{12} \, \phi_2^{(23)}(\tau)
        -
        \frac{23}{22} \, f_{23,1}(\tau)
        - \frac{161}{22} \, f_{23,2}(\tau)
      \right) \,
      \phi_{-2,1}(z;\tau)
      \\[2mm]
      \midrule
      2B &
      2\,{\eta(\tau)^8\over \eta(2\tau)^4}\,\phi_{-2,1}(z;\tau)
      \\[2mm]
      4A &
      2\,{\eta(2\tau)^8\over \eta(4\tau)^4}\,\phi_{-2,1}(z;\tau)
      \\[2mm]
      4C &
      2\,{\eta(\tau)^4 \, \eta(2\tau)^2\over
        \eta(4\tau)^2}\,\phi_{-2,1}(z;\tau)
      \\[2mm]
      3B &
      2\,{\eta(\tau)^6\over \eta(3\tau)^2}\,\phi_{-2,1}(z;\tau)
      \\[2mm]
      6B &
      2\,{\eta(\tau)^2 \, \eta(2\tau)^2 \, \eta(3\tau)^2\over
        \eta(6\tau)^2}\,\phi_{-2,1}(z;\tau)
      \\[2mm]
      12B &
      2\,{\eta(\tau)^4 \, \eta(4\tau) \, \eta(6\tau)\over
        \eta(2\tau) \, \eta(12\tau)}\,\phi_{-2,1}(z;\tau)
      \\[2mm]
      10A &
      2\,{\eta(\tau)^3 \, \eta(2\tau) \, \eta(5\tau)\over
        \eta(10\tau)}\,\phi_{-2,1}(z;\tau)
      \\[2mm]
      12A &
      2\,{\eta(\tau)^3 \, \eta(4\tau)^2 \,\eta(6\tau)^3\over
        \eta(2\tau) \, \eta(3\tau)
        \,\eta(12\tau)^2}\,\phi_{-2,1}(z;\tau)
      \\[2mm]
      21AB &
      \left(
        \frac{7}{3} \,
        \frac{
          \eta(\tau)^3 \,
          \eta(7 \, \tau)^3
        }{
          \eta(3 \, \tau) \, \eta(21 \, \tau)
        }
        - \frac{1}{3} \,
        \frac{ \eta(\tau)^6}{
          \eta(3 \, \tau)^2}   \right) \,
      \phi_{-2,1}(z;\tau)
      \\[2mm]
    \bottomrule
  \end{tabular}
  }
  \caption{Twisted elliptic genus $Z_g(z;\tau)$.}
  \label{tab:twisted_genus}
\end{table}

\subsection{\mathversion{bold}
Twisted Elliptic Genus from the $\eta$-Products of Cycle Shape}


Given a conjugacy class $g$ of $M_{24}$ 
and its cycle shape $1^{r_1}2^{r_2}3^{r_3}\cdots$,
the corresponding $\eta$-product $\eta_g(\tau)$ is defined by 
\begin{equation}
  \eta_g(\tau)
  =
  \prod_i \left[ \eta(i \, \tau) \right]^{r_i} .
\end{equation}  
For instance, 
\begin{alignat*}{3}
  &\eta_{\mathrm{1A}}(\tau)
  =\eta(\tau)^{24},
  &
  \qquad
  &
  \eta_{\mathrm{2A}}(\tau)
  =\eta(\tau)^8\eta(2\tau)^8 ,
  &
  \\
  &
  \eta_{\mathrm{4A}}(\tau)=\eta(\tau)^{4}\eta(2\tau)^2\eta(4\tau)^4,
  &&
  \eta_{\mathrm{8A}}(\tau)=\eta(\tau)^2\eta(2\tau)\eta(4\tau)\eta(8\tau)^2,
  & \qquad
  &
  \cdots
\end{alignat*}

It is well-known that the type~I classes of $M_{24}$ have a balanced shape 
$r_i=r_{{N/i}}$ where $N$ is the order of the element $g$.
As observed in Refs.~\cite{DummKisiMcKa85a,Koike84a,Mason85a},
these $\eta$-products
are cusp forms,
$
\eta_g(\tau) \in
\mathbb{S}_{k}\left(\Gamma_0(N),\chi \right)$,
\emph{i.e.},
weight $k$ cusp form on $\Gamma_0(N)$ with character $\chi$,
and they are the Hecke eigenforms.
See Table~\ref{tab:eta_product} for data of $k$, $N$, and $\chi$.
For our later convention,
we define a Jacobi form
\begin{equation}
  \label{define_varphi}
  \varphi_g(z;\tau) = \eta_g(\tau) \,
  \phi_{-2,1}(z;\tau) ,
\end{equation}
which belongs to $\mathbb{J}_{k-2,1}(\Gamma_0(N),\chi)$,\emph{i.e.},
weight $(k-2)$ Jacobi form on $\Gamma_0(N)$ with index=1 and character
$\chi$.
See Appendix~\ref{sec:modular} for a fundamental property of Jacobi form and
the definition of $\phi_{-2,1}(z;\tau)$.

In order to relate the $\eta$-product to the twisted elliptic genus, we
introduce the Hecke operator.
On the Jacobi form $\phi \in \mathbb{J}_{k,m}(\Gamma_0(N),\chi)$,
the Hecke operator $T_n$ acts as~\cite{EichZagi85}
\begin{equation}
  \left( T_n \phi  \right) (z;\tau)
  =
  n^{k-1}
  \sum_{
    \substack{
      a > 0
      \\
      a d= n
      \\
      (a,N)=1
    }}
  \sum_{b=0}^{d-1}
  \frac{1}{d^k} \,
  \chi(a) \,
  \phi\left(
    a \, z; \frac{a \tau+b}{d}
  \right) .
\end{equation}
The character satisfies $\chi(a)=0$  unless $(a,N)=1$, so we may omit
a condition in the summation.
We note that
$T_n \phi$
belongs to
$\mathbb{J}_{k, m n}\left(\Gamma_0(N),\chi\right)$.

We find that
the twisted elliptic genus $Z_g(z;\tau)$ for type~I conjugacy class
has a simple expression in terms of the
$\eta$-product. Derivation is given in the next section.

For $g=\mathrm{1A}$,
$\mathrm{2A}$,
$\mathrm{3A}$,
$\mathrm{5A}$,
$\mathrm{7AB}$,
$\mathrm{4B}$,
$\mathrm{6A}$, and
$\mathrm{8A}$, 
we have 
\begin{gather}
  Z_{g}(z;\tau)
  =
  - \frac{
    \left(
      T_{2} \varphi_{g}
    \right) (z;\tau)
  }{
    \varphi_{g}(z;\tau)
  } .
  \label{twist_genus_Hecke}
\end{gather}

We need a correction term 
for the remaining type~I cases.
The $\varphi$-functions~\eqref{define_varphi}
for $g=\mathrm{11A}$, $\mathrm{14AB}$, and $\mathrm{15AB}$,
are  weight $0$ modular
forms, and we find
\begin{gather}
  Z_{\mathrm{11A}}(z;\tau)
  =
  - \frac{
    \left(
      T_{2} \varphi_{\mathrm{11A}}
    \right) (z;\tau)
  }{
    \varphi_{\mathrm{11A}}(z;\tau)
  }
  -
  \frac{11}{2} \,
  \varphi_{\mathrm{11A}}(\tau)  ,
  \\
  Z_{\mathrm{14AB}}(z;\tau)
  =
  - \frac{
    \left(
      T_{2} \varphi_{\mathrm{14AB}}
    \right) (z;\tau)
  }{
    \varphi_{\mathrm{14AB}}(z;\tau)
  }
  -
  7 \,
  \varphi_{\mathrm{14AB}} (z;\tau) ,
  \\
  Z_{\mathrm{15AB}}(z;\tau)
  =
  - \frac{
    \left(
      T_{2} \varphi_{\mathrm{15AB}}
    \right) (z;\tau)
  }{
    \varphi_{\mathrm{15AB}}(z;\tau)
  }
  -
  \frac{15}{2} \,
  \varphi_{\mathrm{15AB}}(z;\tau)  .
\end{gather}
For $g=\mathrm{23AB}$ case, the twisted genus can not be expressed only 
in terms of the function $\varphi_{\mathrm{23AB}}$, and we have
\begin{equation}
  Z_{\mathrm{23AB}}(z;\tau)
  =
  - \frac{
    \left(
      T_{2} \varphi_{\mathrm{23AB}} 
    \right) (z;\tau)
  }{
    \varphi_{\mathrm{23AB}}(z;\tau)
  }
  -
  \frac{23}{8}
  \left(
    f_{23,1}(\tau) + 3 \, f_{23,2}(\tau)
  \right)
  \phi_{-2,1}(z;\tau)  .
\end{equation}
See Appendix for new forms of $\Gamma_0(23)$, $f_{23,1}(\tau)$ and $f_{23,2}(\tau)$.

\begin{table}
  \newcolumntype{L}{>{$}l<{$}}
  \newcolumntype{C}{>{$}c<{$}}
  \rowcolors{2}{gray!11}{}
  \centering
  \begin{tabular}{cLCCC}
    \toprule
    $g$ & \eta_g & k &
    N &
    \chi
    \\
    \midrule \midrule
    1A & 1^{24} & 
    12 & 1 &
    \\
    2A & 1^8 \cdot 2^8 & 
    8 & 2 &
    \\
    3A & 1^6 \cdot 3^6 &
    6 & 3 &
    \\
    5A & 1^4 \cdot 5^4&
    4 & 5 &
    \\
    4B &1^4 \cdot 2^2 \cdot 4^4 &
    5 & 4 & \left( \frac{-1}{d} \right)
    \\
    7AB &1^3 \cdot 7^3 &
    3 & 7 & \left( \frac{-7}{d} \right)
    \\
    8A &1^2 \cdot 2^1 \cdot 4^1 \cdot 8^2 &
    3 & 8 & \left( \frac{-2}{d} \right)
    \\
    6A &1^2 \cdot 2^2 \cdot 3^2 \cdot 6^2 &
    4 & 6 &
    \\
    11A&1^2 \cdot 11^2 &
    2 & 11 &
    \\
    15AB &1^1 \cdot 3^1 \cdot 5^1 \cdot 15^1 &
    2 & 15 &
    \\
    14AB &1^1 \cdot 2^1 \cdot 7^1 \cdot 14^1 &
    2 & 14 &
    \\
    23AB &1^1 \cdot 23^1 &
    1 & 23 & \left( \frac{-23}{d} \right)
    \\
    \midrule
    12B &12^2 &
    1 & 144 & \left( \frac{-1}{d} \right)
    \\
    6B &6^4 &
    2 & 36 &
    \\
    4C &4^6 &
    3 & 16 & \left( \frac{-1}{d} \right)
    \\
    3B &3^8 &
    4 & 9 &
    \\
    2B &2^{12} &
    6 & 4 &
    \\
    10A &2^2 \cdot 10^2 &
    2 & 20 &
    \\
    21AB &3^1 \cdot 21^1 &
    1 & 63 & \left( \frac{-7}{d} \right)
    \\
    4A &2^4 \cdot 4^4 &
    4 & 8 &
    \\
    12A &2^1 \cdot 4^1 \cdot 6^1 \cdot 12^1 &
    2 & 24 &
    \\
    \bottomrule
  \end{tabular}
  \caption{$\eta$-product for conjugacy class $g$.
    $\eta_g$ is a weight $k$ modular form
    on $\Gamma_0(N)$ with character $\chi$.
    $\chi$ is written only if non-trivial.
}
  \label{tab:eta_product}
\end{table}


\section{Borcherds Products}
\label{sec:Borcherds}
\subsection{Hilbert Scheme of Points on $K3$}
We first recall the elliptic genus of the Hilbert scheme of  points on the
$K3$ surface~\cite{DijkVerlVerl97b,DijMooVerVer97a}.
We consider a ``second-quantized'' version of $K3$ elliptic surface
\begin{equation}
  \label{partition_symmetric_product}
  \mathsf{M}(\Omega) 
  =
  \sum_{m=0}^\infty Z_{K3}^{\hskip2mm [m]}(z;\tau) \, p^m ,
\end{equation}
where
$p=\E^{2 \pi \I \sigma}$, and
$\Omega=
\left(
  \begin{smallmatrix}
    \tau & z \\
    z & \sigma
  \end{smallmatrix}
\right)
$ is in the Siegel upper half-plane. $Z_{K3}^{\,\,[m]}$ denote the
elliptic genus of the $m$-th symmetric product of $K3$.

As was pointed out in~\cite{DijMooVerVer97a}, 
a generating function $\mathsf{M}(\Omega)$  has an  infinite product
representation
\begin{equation}
  \label{DMVV_product}
  \begin{aligned}[t]
    \mathsf{M}(\Omega)
    & =
    \prod_{m=1}^\infty \prod_{n=0}^\infty \prod_{\ell \in \mathbb{Z}}
    \frac{1}{
      \left(
        1 - p^m \, q^n \, \zeta^\ell
      \right)^{c(n m, \ell)}}\\
    & =
    \exp\left(
      \sum_{m=1}^\infty \left( T_m Z_{K3} \right) ( z;\tau) \, p^m
    \right)
  \end{aligned}
\end{equation}
where $c(n,\ell)$ is the Fourier coefficients of the
$K3$ elliptic genus
\begin{equation}
  \label{Fourier_K3}
  Z_{K3}(z;\tau) = \sum_{n=0}^\infty \sum_{\ell \in \mathbb{Z}}
  c(n,\ell) \, q^n \, \zeta^\ell .
\end{equation}
This construction is based on  the fact that $T_m Z_{K3}$ is a weak
Jacobi form with
weight $0$ and index $m$.
Recalling generators of the Siegel upper half plane given in
Appendix~\ref{sec:Siegel},
we need to symmetrize the infinite product~\eqref{DMVV_product} in
$\tau$ and $\sigma$ to generate 
a Siegel modular form.
After symmetrization and demanding a good modular behavior we obtain
\begin{equation}
  \label{Igusa_form}
  \Phi( \Omega)
  =
  p \, q \, \zeta
  \prod_{
    \substack{
      n, \ell, m \in \mathbb{Z}
      \\
      (n,\ell,m)>0
    }}
  \left( 1 - p^m \, q^n \, \zeta^\ell \right)^{c(n m , \ell)} .
\end{equation}
Here  a condition $(n,\ell,m)>0$ means
\begin{equation*}
  \text{
    $n \geq 0$, 
    $m\geq 0$,
    and
    $
    \begin{cases}
      \ell \in \mathbb{Z}, &
      \text{when $m+n>0$},
      \\
      \ell < 0, &
      \text{when $m,n=0$}.
    \end{cases}
    $
  }
\end{equation*}
The Siegel modular form $\Phi(\Omega)$
is the Igusa cusp form with weight $10$, and equals 
$\mathsf{M}(\Omega)^{-1}$ up to an extra factor 
\begin{equation}
  \label{Hodge_anomaly}
  \begin{aligned}[b]
    \Phi(\Omega) \,
    \mathsf{M}(\Omega)
    & =
    p \, q \, \zeta \, (1- \zeta^{-1})^{2} \,
    \prod_{n=1}^\infty
    (1 - q^n \, \zeta)^2 \,
    (1 - q^n)^{20} \,
    (1 - q^n \, \zeta^{-1})^2 
    \\
    & =
    p \, \varphi_{\mathrm{1A}}(z;\tau) ,
  \end{aligned}
\end{equation}
which is called the Hodge anomaly~\cite{GritsNikul97a,Gritse99a}.

It is well-known that besides being an infinite product,
the Igusa cusp form~$\Phi(\Omega)$ can also be expressed as an infinite sum  
or  Saito--Kurokawa--Maass additive lift,
from a weak Jacobi form with weight $10$, \emph{i.e.}
$\eta(\tau)^{24}\phi_{-2,1}(z;\tau)=\varphi_{1A}(z;\tau)$
\begin{equation}
  \label{Borcherds_additive}
  \Phi(\Omega)=\Phi_{\mathrm{1A}}(\Omega)
  =
  \sum_{m=1}^\infty
  p^m \,
  \left( T_m \varphi_{\mathrm{1A}} \right) (z;\tau) .
\end{equation}
The identity between the infinite product and infinite sum follows from the 
Koecher principle that Siegel modular form with
weight $0$ is a constant.
This relation is interpreted as the analogue of the Weyl denominator
formula for  generalized Kac--Moody
algebra~\cite{Borch95a,GritsNikul97a,GritsNikul98b}.

Our formula of the elliptic genus~\eqref{twist_genus_Hecke} in the
case $g=\mathrm{1A}$ 
\begin{equation*}
  Z_{\mathrm{1A}}(z;\tau)
  =
  -{T_2 \varphi_{\mathrm{1A}}(z;\tau)\over
    \varphi_{\mathrm{1A}}(z;\tau)}
\end{equation*}
simply 
follows from~\eqref{partition_symmetric_product},~\eqref{Hodge_anomaly}
and~\eqref{Borcherds_additive}.

The inverse of the Borcherds product
is a partition function of ${1\over 4}$ 
BPS states, and at a
pole $z=0$ its residue factorizes into a pair of $\frac{1}{2}$ BPS
partition functions~\cite{DijkVerlVerl97b}
\begin{equation}
  \frac{1}{\Phi_{\mathrm{1A}}(\Omega)}
  \underset{z\to 0}{\approx}
  \frac{1}{
    (2 \, \pi \, \I \, z)^2
  }    \,
  \frac{1}{\eta(\tau)^{24}} \,
  \frac{1}{\eta(\sigma)^{24}} 
  =
  {1\over (2 \, \pi  \, \I  \, z)^2} \,
  {1\over \eta_{\mathrm{1A}}(\tau)} \,
  {1\over \eta_{\mathrm{1A}}(\sigma)}.  
\end{equation}

It is to be noted that
the  Hodge anomaly~\eqref{Hodge_anomaly}
is related to the coefficients of BPS characters 
in the character decomposition of the elliptic genus for $K3^{[m]}$,
\begin{equation*}
  \prod_{n=1}^\infty
  \frac{1}{
    (1 - p^n \, \zeta)^2 \,
    (1 - p^n)^{20} \,
    (1 - p^n \, \zeta^{-1})^2
  }
  =
  \sum_{m=0}^\infty p^m \,
  \sum_{s=0}^m
  \gamma_{m,s} \,
  \frac{
    \zeta^{s+1} - \zeta^{-s-1}
  }{
    \zeta - \zeta^{-1}
  } ,
\end{equation*}
where $\gamma_{m,s}$ is the number of the BPS characters of
isospin $\frac{s}{2}$ representation in the elliptic genus of
$K3^{[m]}$~\cite{EguchiHikami08a,EguchiHikami09b}.
See also~\cite{KatKleVaf99a}.


\subsection{Borcherds Product for Twisted Elliptic Genus}
In order to construct an infinite product representation and  
derive~\eqref{twist_genus_Hecke} for general conjugacy classes,  we have to introduce 
another Hecke operator acting on Jacobi forms of the congruence subgroup $\Gamma_0(N)$.
We set the Hecke operator $V_n$ on $\phi\in\mathbb{J}_{k,m}(\Gamma_0(N))$
defined as~\cite{HAokiIbuki05a,ClerGrit11a}
\begin{equation}
  \left(V_n \phi \right)
  (z;\tau)
  =
  n^{k-1}
  \sum_{
    \left(
      \begin{smallmatrix}
        a & b \\
        c & d
      \end{smallmatrix}
    \right)
    \in \Gamma_0(N) \backslash \Mat_n(N)}
  (c \, \tau+d)^{-k} \,
  \E^{-2 \pi \I m n \frac{c z^2}{c\tau + d}} \,
  \phi
  \left(
    \frac{n \,z}{c \, \tau+d}; \frac{a \, \tau+b}{c \, \tau+d}
  \right) ,
\end{equation}
where
\begin{equation*}
  \Mat_n(N) =
  \left\{
    \begin{pmatrix}
      a & b \\
      c  & d
    \end{pmatrix}
    ~\Big|~
    a,b,c,d \in \mathbb{Z},
    a \, d  - b \, c = n,
    c \equiv 0 \bmod N
  \right\}     .
\end{equation*}
We have
$\left( V_n \phi \right) (z;\tau)\in \mathbb{J}_{k,m n}(\Gamma_0(N))$.
The representatives of cosets are given by  cusps of $\Gamma_0(\ord(g))$
as~\cite{HAokiIbuki05a}
\begin{multline}
  \label{representative_V}
  \Gamma_0(N) \backslash \Mat_n(N)
  \\
  =
  \bigsqcup_{
    f/e \in \Cusp(\Gamma_0(N))
  }
  \left\{
    M_{f/e} \,
    \begin{pmatrix}
      a & b \\
      0 & d
    \end{pmatrix}
    ~\Big|~
    a \, d = n,
    a \, e \equiv 0 \bmod N,
    0 \leq b < h_e \,d
  \right\}  ,
\end{multline}
where $\Cusp(\Gamma_0(N))$ denotes the set of cusps of
$\Gamma_0(N)$, and
$
M_{f/e} =
\left(
  \begin{smallmatrix}
    f & * \\
    e & *
  \end{smallmatrix}
\right)
\in SL(2;\mathbb{Z})  
$ with a positive divisor $e$ of $N$.
$h_e$ is the width of cusp $f/e$,
\begin{equation*}
  h_e= \frac{N}{(e^2,N)} .
\end{equation*}
When $N$ is prime, the Hecke subgroup $\Gamma_0(N)$ has two cusps,
$\tau=\I \infty $ and $0$.
The width is $1$ and $N$, respectively.

By use of $Z_g(z;\tau) \in \mathbb{J}_{0,1}(\Gamma_0(\ord(g)))$,
we introduce  an analogue of~\eqref{DMVV_product} 
\begin{equation}
\label{analogue_DMVV_product}
  \mathsf{M}_g(\Omega)
  =\exp \left[
  \sum_{m=1}^\infty p^m \,
  \left( V_m    Z_g \right) (z;\tau)\right]  .
\end{equation}
In order to rewrite this into an infinite product form,
we introduce the Fourier 
expansion of $Z_g(z;\tau)$ at each cusp $f/e$ of $\Gamma_0(\ord(g))$,
\begin{equation}
  \label{Fourier_other_cusps}
  \left( Z_g |_{0,1} M_{f/e} \right)(z;\tau)
  =
  \sum_{n \in \mathbb{Z}/h_e} \sum_{\ell \in \mathbb{Z}}
  c_{g, f/e}(n,\ell) \, q^n \, \zeta^\ell .
\end{equation}
(For the definition of slash operator, see Appendix~\ref{sec:modular}).
Note that the Fourier expansion at a cusp is a power series in  
$q^{\frac{1}{h_e}}$.

It is well-known that 
$c_{g,f/e}(n,\ell)$ depends only on $4 n-\ell^2$~\cite{EichZagi85} and we may 
write 
\begin{equation}
c_{g,f/e}(n,\ell) 
=  c_{g,f/e}(4 \,n - \ell^2).
\end{equation}
We have
\begin{align*}
  & \log \mathsf{M}_g(\Omega)
  \\
  & =
  \sum_{m=1}^\infty
  \frac{1}{m} \, p^m
  \sum_{f/e\in \Cusp(\Gamma_0(\ord(g)))}
  \sum_{\substack{
      ad=m \\
      ae= 0 \bmod \ord(g)
    }}
  \sum_{b=0}^{h_e d-1} \sum_{n \in \mathbb{Z}/h_e}
  \sum_\ell
  c_{g,f/e}(n,\ell) \,
  q^{n \frac{a}{d}} \, \zeta^{\ell m/d} \,
  \E^{2 \pi \I n b/d}
  \\
  & =
  \sum_{f/e\in \Cusp(\Gamma_0(\ord(g)))}
  \sum_n \sum_\ell
  \sum_d 
  \frac{h_e}{N_e} \, c_{g,f/e}(n \, d, \ell) 
  \sum_{a^\prime=1}^\infty
  \frac{
    1}{a^\prime} \,
  \left( p^d \, q^n \, \zeta^\ell \right)^{N_e a^\prime} ,
\end{align*}
where we have used
\begin{equation}
N_e=\frac{\ord(g)}{e}.
\end{equation}
As a result, we obtain~\cite{HAokiIbuki05a,ClerGrit11a}
\begin{equation}
  \label{portion_twist_Borcherds}
  \mathsf{M}_g(\Omega)
  =
  \prod_{
    f/e \in \Cusp(\Gamma_0(\ord(g)))
  }
  \prod_{m=1}^\infty
  \prod_{n=0}^\infty \prod_{\ell \in \mathbb{Z}}
  \frac{1}{
    \left(
      1 -
      \left(p^m \, q^n \, \zeta^\ell \right)^{N_e}
    \right)^{
      \frac{h_e}{N_e} c_{g, f/e}(m n, \ell)
    }
  }.
\end{equation}
For the contribution of the cusp at $\I\infty$ 
we set 
$c_{g,\I \infty}(n,\ell)=c_g(n,\ell)$\
which is the Fourier expansion coefficients of the elliptic genus $Z_g$ and 
$N_e=h_e=1$.

It should be noted that 
only
the Fourier coefficients at  integral powers
in~\eqref{Fourier_other_cusps} contribute to the infinite
product~\eqref{portion_twist_Borcherds}.
In  Table~\ref{tab:Fourier1} we have tabulated  
values of the Fourier
coefficients
$c_{g,f/e}(m)$ for integers $m\leq 32$.


By inspection we notice an interesting relation 
\begin{equation}\label{sum_rule}
  \sum_{
    f/e \in \Cusp(\Gamma_0(\ord(g)))
  } h_e \, c_{g,f/e}(n)
  = c_{\mathrm{1A}}(n)
\end{equation}
for all $g \in \text{type~I}$.
Namely, sum of the expansion
coefficients for each  cusp weighted by the width reproduces the
original Fourier coefficients of $K3$ elliptic genus.\footnote{
  There exist more linear relations among Fourier coefficients
  such as (we abbreviate $c_{g,\I\infty}$ to $c_g$)
  \begin{gather*}
    c_{\mathrm{2A}} - c_{\mathrm{4B}}
    =c_{\mathrm{4B},{1\over 2}},
    \qquad
%
    c_{\mathrm{3A}}-c_{\mathrm{6A}}
    =2c_{\mathrm{6A},{1\over 3}},
    \qquad
%
    c_{\mathrm{1A}}-c_{\mathrm{2A}}=
    8c_{\mathrm{8A},0},
    \\
    c_{\mathrm{7AB}}-c_{\mathrm{14AB}}
    =2c_{\mathrm{14AB},{1\over 7}},
    \qquad
%
    c_{\mathrm{3A}}-c_{\mathrm{15AB}}
    =5c_{\mathrm{15AB},{1\over 3}},
    \qquad
    \text{etc.} 
  \end{gather*}
  Making use of these relations and~\eqref{sum_rule} one can derive a
  different expression for  $\mathsf{M}_g(\Omega)$ for all
  $g\in  \text{type~I}$
  \begin{equation*}
    \mathsf{M}_g(\Omega)=
    \exp
    \left[\sum_{m,n,\ell}\sum_{k=1}^\infty
      {1\over   k} \, c_{g^k}(4mn-\ell^2) \,
      \left(p^mq^n\zeta^{\ell}\right)^k
    \right] ,
  \end{equation*}
  which is used in some literature (see,
  \emph{e.g.},~\cite{GovinKrish09b,MCheng10a,ClerGrit11a}).}

Using the Table 4 we note that the $\eta$ product for all type~I class
$g$ can be uniformly written as 
\begin{equation}\label{eta_general}
  \eta_g(\tau)=\prod_{
    f/e \in \Cusp(\Gamma_0(\ord(g)))
  }
  \eta(N_e\tau)^{{h_e\over  N_e}\left[c_{g,f/e}(0)+2c_{g,f/e}(-1)\right]} .
\end{equation}
Here we see how the cycle decomposition of an element $g$ corresponds
to the decomposition into cusps of $\Gamma_0(\ord(g))$.
Length of each
cycle equals  $N_e$ and its power is 
given by ${h_e\over N_e}\left[c_{g,f/e}(0)+2c_{g,f/e}(-1)\right]$.

As before, by symmetrizing $p$ and $q$ in~\eqref{portion_twist_Borcherds}
we set \cite{ClerGrit11a}
\begin{multline}
  \label{define_twist_Borcherds}
  \Phi_g(\Omega)
  =
  p^{\frac{1}{4} \sum_\ell \ell^2 c_{g}(0, \ell)} \,
  q^{\frac{1}{24} \sum_\ell c_g(0,\ell)} \,
  \zeta^{\frac{1}{2} \sum_{\ell >0} \ell c_g(0,\ell)}
  \\
  \times
  \prod_{
    f/e \in \Cusp(\Gamma_0(\ord(g)))
  }
  \prod_{
    \substack{
      n, m, \ell \in \mathbb{Z}
      \\
      (n,m,\ell)>0
    }}
  \left(
    1 - 
    \left( p^m \, q^n \, \zeta^\ell \right)^{N_e}
  \right)^{\frac{h_e}{N_e} c_{g,f/e}(n m, \ell)} .
\end{multline}
By checking transformation properties under generators of a congruence
subgroup given in Appendix~\ref{sec:Siegel},
we  see that $\Phi_g(\Omega)$
is the Siegel modular form on
$\Gamma_0^{(2)}(\ord(g))$~\cite{HAokiIbuki05a,ClerGrit11a}.
Comparing with~\eqref{portion_twist_Borcherds}, we find the Hodge
anomaly
\begin{equation}
  \label{twist_Borcherds_varphi}
  \Phi_g(\Omega) \, \mathsf{M}_g(\Omega)
  =
  p \, \varphi_g(z;\tau) .
\end{equation}

Furthermore
as shown in~\cite{ClerGrit11a}, the Borcherds product can be written
as an
additive lift of the $\eta$-product
\begin{equation}
  \label{twist_Borcherds_additive}
  \Phi_g(\Omega)
  =
  \sum_{m=1}^\infty
  p^m \,
  \left( T_m \varphi_g\right)
  (z;\tau)  ,
\end{equation}
for $g=\mathrm{1A}$,
$\mathrm{2A}$,
$\mathrm{3A}$,
$\mathrm{5A}$,
$\mathrm{7AB}$,
$\mathrm{4B}$,
$\mathrm{6A}$, and
$\mathrm{8A}$.
Our formula~\eqref{twist_genus_Hecke}
follows
from~\eqref{analogue_DMVV_product},~\eqref{twist_Borcherds_varphi},
and~\eqref{twist_Borcherds_additive}.

In the case of remaining conjugacy classes, the weight of Jacobi form 
$\varphi_g$ becomes zero or negative 
and the situation is somewhat complex.
For $g=\mathrm{11A}$, we conjecture the following relation for the Borcherds
product~\eqref{define_twist_Borcherds}
\begin{equation}
  \Phi_{\mathrm{11A}}(\Omega)
  =
  \frac{1}{11} \, 
  \left[
    \exp\left(
      11
      \sum_{m=1}^\infty p^m \, 
      \left(
        T_m \varphi_{11A}
      \right)  (z;\tau)
    \right)
    -1
  \right] .
\end{equation}
Similar relations are conjectured for $\mathrm{14AB}$ and $\mathrm{15AB}$,
\begin{gather}
  \Phi_{\mathrm{14AB}}(\Omega)
  =
  \frac{1}{14} \, 
  \left[
    \exp\left(
      14
      \sum_{m=1}^\infty p^m \, 
      \left(
        T_m \varphi_{14AB}
      \right)  (z;\tau)
    \right)
    -1
  \right] ,
  \\
  \Phi_{\mathrm{15AB}}(\Omega)
  =
  \frac{1}{15} \, 
  \left[
    \exp\left(
      15
      \sum_{m=1}^\infty p^m \, 
      \left(
        T_m \varphi_{15AB}
      \right)  (z;\tau)
    \right)
    -1
  \right] .
\end{gather}
We do not have an analogous expression for 
$\mathrm{23AB}$.

We note that exactly the same exponent as~\eqref{eta_general} appears in the RHS
of~\eqref{define_twist_Borcherds}.
Then again the inverse of $\Phi_g(\Omega)$ factorizes at its pole into
$\eta$ products,  and we obtain 
\begin{equation}
  \frac{1}{
    \Phi_g(\Omega)
  }
  \underset{z\to 0}{\approx}
  \frac{1}{(2 \, \pi \, \I \, z)^2} \,
  \frac{1}{
    \eta_g(\tau)
  } \,
  \frac{1}{\eta_g(\sigma)} ,
\end{equation}
for all $g \in \text{type~I}$.

\begin{table}
  \newcolumntype{R}{>{$}r<{$}}
  \newcolumntype{C}{>{$}c<{$}}
  \rowcolors{2}{gray!11}{}
  \centering
  \rotatebox[]{90}{
  \resizebox{0.93\textheight}{!}{
    \centering
    \begin{tabular}{cCCC||RRRRRRRRRRRRRRRRRR}
      \toprule
      $g$& \text{cusp} & h_e & N_e &
      -1 & 0 & 3 & 4 & 7 & 8 &11 & 12    &
      15 & 16 & 19 & 20 & 23 & 24 & 27 & 28 &31 & 32
      \\
      \midrule \midrule
      1A & \I \infty & 1 & 1&
      2 & 20 & -128 & 216 & -1026 & 1616 & -5504 & 8032 & -23550 &
      33048 &
      -86400 & 117280 & -283652 & 376608 & -854528 & 1112832 
      & -2402298 & 3082192
      \\
      \midrule
      2A & \I \infty & 1 & 1 &
      2 & 4 & 0 & -8 & -2 & 16 & 0 & -32 &     2 & 56 & 0 & -96 &
      -4 & 160 & 0 & -256& 6 & 400
      \\
      & 0 & 2 & 2 &
      0 & 8 &-64  &112 & -512 & 800 & -2752 & 4032 &      -11776 &
      16496 & 
      -43200 &58688 & -141824 &188224 & -427264 & 556544 &
      -1201152 & 1540896
      \\
      \midrule
      3A & \I \infty & 1 & 1 &
      2 & 2 & -2 & 0 & 0 & -4 & 4 & 4 & -6 & 0 & 0 & -8 &
      10 & 12 & -14 & 0 & 0 & -20
      \\
      & 0 & 3 & 3 &
      0 & 6 & -42 & 72 & -342 &540 & -1836 & 2676 & -7848 & 11016 &
      -28800 & 39096 & -94554 & 125532 & -284838 & 370944 &
       -800766 & 1027404
      \\
      \midrule
      4B & \I \infty & 1&  1 &
      2& 0 & 0 & 0& -2 & 0 & 0 & 0 & 2 & 0 &0 & 0 &
      -4 & 0  & 0 & 0   & 6 & 0
      \\
      & 0 & 4 & 4 &
      0 & 4 & -32 & 56 & -256 & 400 & -1376 & 2016 &
      -5888& 8248 & -21600 & 29344 & -70912 & 94112 &
      -213632 & 278272 & -600576 &770448
      \\
      & \frac{1}{2} & 1 &  2 &
      0 & 4 & 0 & -8 & 0 & 16 & 0 & -32 &
      0 & 56 &  0 & -96 & 0 & 160 & 0 & -256 & 0 & 400
      \\
      \midrule
      5A & \I \infty  & 1 &  1 &
      2 & 0 & 2 & -4 & 4 & -4 & 6 & -8 & 10 & -12& 20 & -20 & 28 &
      -32& 42 & -48 & 62 & -68
      \\
      & 0 & 5 & 5 &
      0  & 4 & -26 & 44 & -206 & 324 & -1102 & 1608 & -4712 & 6612 &
      -17284 & 23460 & -56736 & 75328 & -170914 & 222576 & -480472 &
      616452
      \\
      \midrule
      6A & \I \infty & 1 & 1 &
      2 & -2 & 6 & -8 & 16 & -20 & 36 &-44 & 74 & -88 &
      144 & -168 & 266 & -308& 474& -544 & 816 & -932
      \\
      & 0 & 6 & 6 &
      0 & 2 & -20 & 36 & -168& 264 & -912& 1336& -3912 & 5484 &
      -14376 & 19536 & -47232 & 62688 & -142340 & 185424 &
      -400248 & 513480
      \\
      & \frac{1}{2} & 3 &  3 &
      0 & 2  & -2 & 0 & -6 & 12 & -12 & 4 & -24 & 48 & -48 & 24 &
      -90 & 156 & -158 & 96 & -270 &444
      \\
      & \frac{1}{3} & 2 & 2 &
      0 & 2  & -4 & 4 & -8 & 8 & -16 &24 & -40 & 44 & -72 &80 &
      -128 & 160 & -244 & 272 & -408 & 456
      \\
      \midrule
      7AB & \I \infty & 1 & 1 &
      2 & -1 & 5 & -8 & 17 & -22& 47 &-60 & 110 & -132 &
      239 & -292 & 492 & -580 & 963 & -1134 & 1810 & -2106
      \\
      & 0 & 7 & 7 &
      0 & 3 & -19 & 32 & -149 & 234 & -793 & 1156 & -3380 & 4740 &
      -12377 & 16796 & -40592 & 53884 & -122213 & 159138 & -343444&
      440614
      \\
      \midrule
      8A & \I \infty & 1 & 1 &
      2 & -2 &8 & -12 & 30 & -40 & 88 & -112 & 226 & -284 &
      536 & -656    &
      1180 & -1424 & 2464 & -2944 & 4934 & -5832
      \\
      & 0 & 8 & 8 &
      0 & 2 & -16 & 28 & -128 & 200 & -688& 1008& -2944 & 4124 &
      -10800 & 14672 &
      -35456 & 47056 & -106816 & 139136 & -300288 & 385224
      \\
      & \frac{1}{2} & 2 & 4 &
      0 & 2 & 0 & -4 & 0 & 8 & 0 & -16 & 0 & 28 & 0 & -48 & 0 & 80 &
      0 & -128 & 0 & 200
      \\
      & \frac{1}{4} & 1 & 2 &
      0 & 2 & -8 & 12 & -32 & 40 & -88& 112 & -224 & 284 & -536 & 656
      &
      -1184 & 1424 & -2464 & 2944 & -4928 & 5832
      \\
      \midrule
      11A & \I \infty & 1 & 1 &
      2 & -2 & 4 & -4 & 8 & -12 & 18 & -20 & 34 & -40 &
      60 & -68 & 104 & -120 & 172 & -192 & 278 & -316
      \\
      & 0 & 11 & 11 &
      0 & 2 & -12 & 20 & -94 & 148 & -502 & 732 & -2144 & 3008 & -7860
      & 10668 & -25796 & 34248 & -77700 & 101184 & -218416 & 280228
      \\
      \midrule
      14AB & \I \infty & 1 & 1 &
      2 & -3 & 7 & -8 & 19 & -26 &49 & -60 & 114 & -140 &
      245 & -292& 500 & -596 & 973 & -1138 & 1826 & -2134
      \\
      &0 & 14 &  14 &
      0 & 1 & -9 & 16 & -73 & 114 & -393 & 576 &
      -1682 & 2356 & -6171 & 8384 & -20260 & 26888 & -61037 & 79506
      & -171592 & 220126
      \\
      & \frac{1}{2} & 7 & 7 &
      0 & 1 & -1 & 0 & -3 & 6 & -7 & 4 & -16 & 28 &
      -35 & 28 & -72 & 108 & -139 & 126 & -260 & 362
      \\
      & \frac{1}{7} & 2 & 2 &
      0 & 1 & -1 & 0 & -1 & 2 & -1 & 0 & -2 & 4 &
      -3 & 0 & -4 & 8  & -5 & 2 & -8 & 14
      \\
      \midrule
      15AB & \I \infty & 1 & 1 &
      2 & -3 & 8 & -10 & 25 & -34& 69 & -86 & 169 & -210 & 380 &
      -458    &      805 & -968 & 1626 & -1920 & 3155 & -3710
      \\
      & 0 & 15 & 15 &
      0 & 1 & -8 & 14 & -67 & 106 & -363 & 530 &
      -1559& 2190 & -5736 & 7790 & -18859 & 25044 &-56862 & 74064 &
      -159947 & 205238
      \\
      & \frac{1}{3} & 5 &  5 &
      0 & 1 & -2 & 2 & -5 & 6 & -13 & 18 & -35 & 42 & -786 & 90      &
      -159 & 196 & -328 & 384 & -631 & 738
      \\
      & \frac{1}{5} & 3 & 3 &
      0 & 1 & -2 & 2 & -7 & 10 & -21 & 26 & -53 & 66 & -120 & 146 &
      -259 & 312 & -528 & 624 & -1031 & 1214
      \\
      \midrule
      23AB & \I \infty & 1 & 1 &
      2 & -3 & 10 & -14 & 32 & -40 & 85 & -110 & 209 & -256 &
      471 & -572 & 996 & -1190 & 2015 & -2392 & 3916 & -4592
      \\
      & 0 & 23 & 23 &
      0 & 1 & -6 & 10 & -46 & 72 & -243 & 354 & -1033 & 1448 & -3777&
      5124 & -12376 & 16426 & -37241 & 48488 & -104618 & 134208
      \\
      \bottomrule
    \end{tabular}
  }
  }
  \caption{The Fourier coefficient
    $c_{g,f/e}(4n-\ell^2)=c_{g,f/e}(n,\ell)
    $
    of the twisted elliptic genus at cusp $f/e$ of $\Gamma_0(\ord(g))$.
  }
  \label{tab:Fourier1}
\end{table}

\section{Concluding Remarks}
In this paper we tried to obtain a simple and 
direct relationship between $\eta$-products of 
various conjugacy classes of $M_{24}$ and the corresponding twisted
elliptic genus of $K3$ surface.
It seems that the simplest way to
derive such a relation is to use the 
identity of Siegel modular forms which may be 
constructed either from Borcherds products or 
Saito--Kurokawa additive lifts of Jacobi forms. 

Relationship (\ref{twist_genus_Hecke}) seems to exhibit some deep
relation between $M_{24}$ and $K3$ surface.
RHS is based purely on
$M_{24}$ and has nothing to do with $K3$.
It, 
however, 
coincides with LHS which is the twisted genus of $K3$.    

We have so far discussed (\ref{twist_genus_Hecke}) and its variation
only in the case of type~I classes.
If one tries to find a similar relation for type~II classes, one obtains 
\begin{gather*}
  Z_{\mathrm{2B}}(z;\tau)
  =
  - \frac{
    \left(
      T_{2} \varphi_{\mathrm{2B}} 
    \right) (z;\tau)
  }{
    \varphi_{\mathrm{2B}}(z;\tau)
  } ,
  \\
  Z_{\mathrm{4A}}(z;\tau)
  =
  - \frac{
    \left(
      T_{2} \varphi_{\mathrm{4A}}
    \right) (z;\tau)
  }{
    \varphi_{\mathrm{4A}}(z;\tau)
  } ,
  \\
  Z_{\mathrm{10A}}(z;\tau)
  =
  - \frac{
    \left(
      T_{2} \varphi_{\mathrm{10A}}
    \right) (z;\tau)
  }{
    \varphi_{\mathrm{10A}}(z;\tau)
  }
  - 10 \, \varphi_{\mathrm{10A}}(z;\tau)  ,
  \\
  Z_{\mathrm{12A}}(z;\tau)
  =
  - \frac{
    \left(
      T_{2} \varphi_{\mathrm{12A}}
    \right) (z;\tau)
  }{
    \varphi_{\mathrm{12A}}(z;\tau)
  }
  -12 \, \varphi_{\mathrm{12A}}(z;\tau)  ,
\end{gather*}
which has the same form as the type~I classes.
 
Unfortunately, in the case of other type~II classes 
we obtain 
expressions which do not seem to clarify the relationship between 
$M_{24}$ and $K3$ surface
\begin{gather*}
  Z_{\mathrm{4C}}(z;\tau)
  =
  - \frac{
    \left(
      T_{2} \varphi_{\mathrm{4C}}
    \right) (z;\tau)
  }{
    \varphi_{\mathrm{4C}}(z;\tau) 
  }
  - 16 \,
  \frac{
    \eta(2 \tau)^4 \, \eta( 8 \tau)^4
  }{
    \eta(4 \tau)^4
  } \, \phi_{-2,1}(z;\tau) ,
  \\
  Z_{\mathrm{3B}}(z;\tau)
  =
  - \frac{
    \left(
      T_{2} \varphi_{\mathrm{3B}}
    \right) (z;\tau)
  }{
    \varphi_{\mathrm{3B}}(z;\tau)
  }
  - 18 \,
  \frac{
     \eta( \tau)^3 \, \eta( 9 \tau)^3
  }{
    \eta(3 \tau)^2
  } \, \phi_{-2,1}(z;\tau) ,
  \\
  \begin{aligned}
    Z_{\mathrm{6B}}(z;\tau)
    & =
    - \frac{
      \left(
        T_{2} \varphi_{\mathrm{6B}}
      \right) (z;\tau)
    }{
      \varphi_{\mathrm{6B}}(z;\tau) 
    }
    \\
    & 
    -2 \,
    \left(
      \eta_{\mathrm{6B}}(\tau)
      +
      \frac{
        \eta(\tau)^3 \, \eta(9\tau)^3
      }{
        \eta(3 \tau)^2
      }
      + 6
      \frac{
        \eta(2\tau)^3 \, \eta(18\tau)^3
      }{
        \eta(6 \tau)^2
      }
      + 8
      \frac{
        \eta(4\tau)^3 \, \eta(36\tau)^3
      }{
        \eta(12 \tau)^2
      }
    \right) \, \phi_{-2,1}(z;\tau) ,\\
  \end{aligned}
  \\
  \cdots   \cdots.
\end{gather*}
As a whole,  type~I classes are reasonably under good control while 
we still know very little about type~II classes.
Twisted genera of type~II classes
have vanishing Witten index and appear to have a little contact with
the classical geometry of $K3$ surface.
On the other hand its character expansion is 
described in terms of  modular forms and should be easier to handle
than the type~I  classes.
We hope to report progress on these issue in the near future.

\section*{Acknowledgments}
T.E. would like thank California Institute for Technology
and profs. H.~Ooguri and J.H.~Schwarz for  
Moore distinguished scholarship during the fall of 2011 and kind   
hospitality.
K.H. thanks the Simons Center for Geometry and Physics for hospitality
in the summer of 2011.
Authors would like to thank H.~Aoki for sending them an unpublished
manuscript.
Thanks are also to M.~Kaneko for communications.
\\
This work is supported in part by Grant-in-Aid from the Ministry of
Education, Culture, Sports, Science and Technology of Japan.

\newpage
\appendix
\section{Modular Form}
\label{sec:modular}
We collect facts about  modular forms.
See, \emph{e.g.},~\cite{BGHZ08Book,EichZagi85}, for details.

\subsection{\mathversion{bold}Jacobi Theta Functions and Dedekind
  $\eta$-function}
The Jacobi theta functions are defined by
\begin{equation}
  \begin{aligned}
    \theta_{11}(z;\tau)
    & =
    \sum_{n \in \mathbb{Z}}
    q^{\frac{1}{2} \left( n+ \frac{1}{2} \right)^2} \,
    \E^{2 \pi \I \left(n+\frac{1}{2} \right) \,
      \left( z+\frac{1}{2} \right)
    }
    ,
    \\[2mm]
    \theta_{10}(z;\tau)
    & =
    \sum_{n \in \mathbb{Z}}
    q^{\frac{1}{2} \left( n + \frac{1}{2} \right)^2} \,
    \E^{2 \pi \I \left( n+\frac{1}{2} \right) z}
    ,
    \\[2mm]
    \theta_{00} (z;\tau)
    & =
    \sum_{n \in \mathbb{Z}}
    q^{\frac{1}{2} n^2} \,
    \E^{2 \pi \I  n  z}
    ,
    \\[2mm]
    \theta_{01} (z;\tau)
    & =
    \sum_{n \in \mathbb{Z}}
    q^{\frac{1}{2} n^2} \,
    \E^{2 \pi \I n \left( z+\frac{1}{2} \right) }
    .
  \end{aligned}
\end{equation}

The Dedekind $\eta$-function is 
\begin{equation}
  \eta(\tau) =
  q^{\frac{1}{24}}
  \prod_{n=1}^\infty
  (1-q^n) .
\end{equation}
We set the Eisenstein series by
\begin{equation}
  \phi_2^{(N)}(\tau)
  =
  \frac{24}{N-1} \, q \, \frac{\partial}{\partial q}
  \log \left(
    \frac{\eta(N  \tau)}{\eta(\tau)}
  \right) .
\end{equation}

We  introduce  new forms of level-$23$
\begin{equation}
  \begin{gathered}[b]
    f_{23,1}(\tau)
    =
    \frac{1}{4} \, \left[ \Theta_1(\tau) \right]^2
    +\frac{1}{2} \, \Theta_1(\tau) \, \Theta_2(\tau)
    - \frac{3}{4} \, \left[ \Theta_2(\tau) \right]^2 ,
    \\[2mm]
    \begin{aligned}
      f_{23,2}(\tau)
      & =
      \left[
        \eta(\tau) \, \eta(23  \tau)
      \right]^2
      \\
      & =
    \frac{1}{4} \, \left[ \Theta_1(\tau) \right]^2
    - \frac{1}{2} \, \Theta_1(\tau) \, \Theta_2(\tau)
    + \frac{1}{4} \, \left[ \Theta_2(\tau) \right]^2 ,
  \end{aligned}
\end{gathered}
\end{equation}
where the $\Theta$-functions are
\begin{equation*}
  \begin{gathered}[b]
    \Theta_1(\tau)
    =
    \sum_{m,n \in \mathbb{Z}}
    q^{m^2+m n + 6n^2} ,
    \\[2mm]
    \Theta_2(\tau)
    =
    \sum_{m,n \in \mathbb{Z}}
    q^{2 m^2+ m n + 3 n^2} .
  \end{gathered}
\end{equation*}
\subsection{Modular Form}
We set the slash operator
\begin{equation}
  \left( f |_k \gamma \right) (\tau)
  =
  \left( \det \gamma \right)^{\frac{k}{2}} \,
  \left( c\, \tau + d \right)^{-k} \,
  f(\gamma \, \tau) ,
\end{equation}
where
$\gamma=
\left(
  \begin{smallmatrix}
    a & b \\
    c & d
  \end{smallmatrix}
\right)$ is an integral matrix.
A modular form $f(\tau) \in \mathbb{M}_k(\Gamma_0(N),\chi)$ satisfies
\begin{equation}
  \left( f |_k \gamma \right)
  (\tau)
  =
  \chi(d) \, f(\tau)
\end{equation}
for
$\gamma=\left(
  \begin{smallmatrix}
    a & b \\
    c & d
  \end{smallmatrix}
\right)
 \in \Gamma_0(N)$,
and  $\chi$ is a Dirichlet character modulo $N$.

\subsection{Jacobi Form}

We define the slash operator
\begin{equation}
  \left( \phi |_{k,m} \gamma \right)
  (z;\tau)
  =
  \left(\det \gamma \right)^{\frac{k}{2}} \,
  \left( c \, \tau + d \right)^{-k} \,
  \E^{2 \pi \I m \frac{ - c z^2}{c \tau+d} } \,
  \phi
  \left(
    \frac{z}{c \, \tau+d } ; \frac{a \, \tau+b}{c \, \tau+d} 
  \right) ,
\end{equation}
for an integral matrix
$\gamma=
\left(
  \begin{smallmatrix}
    a & b \\
    c & d
  \end{smallmatrix}
\right)$.
Then
the Jacobi form $\phi(z;\tau)$
with weight $k$ and index $m \in \mathbb{Z}$ 
on $\Gamma_0(N)$ with $\chi(d)$ fulfills
\begin{equation}
  \begin{aligned}
    \left( \phi |_{k,m} \gamma \right)(z;\tau)
    & =
    \chi(d) \,
    \phi(z;\tau),
    \\[2mm]
    \phi(z + s \, \tau + t; \tau)
    & =
    \E^{-2 \pi \I m (s^2 \tau+ 2 s z)} \,
    \phi(z;\tau) ,
  \end{aligned}
\end{equation}
where
$\gamma =
\left(
  \begin{smallmatrix}
    a & b \\
    c & d
  \end{smallmatrix}
\right)
\in \Gamma_0(N)$
and $s, t \in \mathbb{Z}$.
We set
such space as
$\mathbb{J}_{k,m}(\Gamma_0(N), \chi)$.
Examples of the Jacobi forms $
\mathbb{J}_{k,m}(\Gamma(1))$ are as follows~\cite{EichZagi85};
\begin{align}
  \phi_{-2,1}(z;\tau)
  &  =
  -
  \frac{\left[ \theta_{11}(z;\tau) \right]^2}{
    \left[ \eta(\tau) \right]^6
  } ,
  \label{def_phi-21}
  \\[2mm]
  \phi_{0,1}(z;\tau)
  & =
  4 \, \left[
    \left(
      \frac{\theta_{10}(z;\tau)}{\theta_{10}(0;\tau)}
    \right)^2
    +
    \left(
      \frac{\theta_{00}(z;\tau)}{\theta_{00}(0;\tau)}
    \right)^2
    +
    \left(
      \frac{\theta_{01}(z;\tau)}{\theta_{01}(0;\tau)}
    \right)^2
  \right] .
\end{align}


\subsection{Siegel Modular Form}
\label{sec:Siegel}
The Siegel modular form of degree-$2$ and weight $k$ is a function of
$\Omega=
\left(
  \begin{smallmatrix}
    \tau & z \\
    z & \sigma
  \end{smallmatrix}
\right)
$ in the Siegel upper half plane satisfying
\begin{equation}
  F
  \left(
    (A \, \Omega + B) \,
    ( C \, \Omega + D)^{-1}
  \right)
  =
  \det(C \, \Omega + D)^{k}  \,
  F(\Omega) . 
\end{equation}
Here
$M=
\left(
  \begin{smallmatrix}
    A & B\\
    C & D
  \end{smallmatrix}
\right)
\in Sp(2;\mathbb{Z})
$ fulfills
$
M^{\mathrm{T}} \, J \, M = J
$ for
$
J=
\left(
  \begin{smallmatrix}
    0 & 1_2 \\
    -1_2 & 0
  \end{smallmatrix}
\right)
$.
We use the congruence subgroup $\Gamma_0^{(2)}(N)$ given by
\begin{equation*}
  \Gamma_0^{(2)}(N)
  =
  \left\{
    \begin{pmatrix}
      A & B \\
      C & D
    \end{pmatrix}
    \in Sp(2;\mathbb{Z})
    ~\Big|~
    C = 0 \bmod N
  \right\} ,
\end{equation*}
Its generators are~\cite{HAokiIbuki05a}
\begin{alignat*}{4}
  &
  \begin{pmatrix}
    0 & 1 & 0 & 0 \\
    1 & 0 & 0 & 0 \\
    0 & 0 & 0 & 1 \\
    0 & 0 & 1 & 0
  \end{pmatrix} ,
  & \quad 
  &
  \begin{pmatrix}
    1 & 0 & 0 & 0 \\
    n & 1 & 0 & 0 \\
    0 & 0 & 1 & -n \\
    0 & 0 & 0 & 1
  \end{pmatrix} ,
  & \quad
  &
  \begin{pmatrix}
    a & 0 & b & 0 \\
    0 & 1 & 0 & 0 \\
    c \, N & 0 & d & 0 \\
    0 & 0 & 0 & 1
  \end{pmatrix},
  & \quad
  &
  \begin{pmatrix}
    1_2 & B \\
    0 & 1_2
  \end{pmatrix} ,
\end{alignat*}
where
$n\in \mathbb{Z}$,
$\left(
  \begin{smallmatrix}
    a & b \\
    c N & d
  \end{smallmatrix}
\right)
\in \Gamma_0(N)
$, and
$B^{\mathrm{T}}=B$.
Note that the first one corresponds to symmetrization of $\tau$ and $\sigma$.

\section{\mathversion{bold}$\mathcal{N}=4$ Superconformal Characters}
The characters are defined by
\begin{equation}
  \ch^{\widetilde{R}}_{h,\ell}(z;\tau)
  =
  \Tr_{R}
  \left(
    (-1)^F
    \E^{
      4 \pi \I z T^3_0} \, q^{L_0 - \frac{c}{24}}
  \right) ,
\end{equation}
where $R$ means the Ramond sector of the theory.
They are explicitly given as follows~\cite{EgucTaor86a,EgucTaor88a};
\begin{itemize}
\item BPS (massless) representations ($h=\frac{1}{4}$)
  \begin{gather}
    \ch^{\widetilde{R}}_{h=\frac{1}{4}, \ell=0}(z;\tau)
    =
    \frac{
      \left[ \theta_{11}(z;\tau) \right]^2}{
      \left[ \eta(\tau) \right]^3} \,
    \mu(z;\tau),
    \\
    \ch^{\widetilde{R}}_{h=\frac{1}{4},\ell=\frac{1}{2}}(z;\tau)
    + 2 \,
    \ch^{\widetilde{R}}_{h=\frac{1}{4},\ell=0}(z;\tau)
    =
    q^{-\frac{1}{8}} \, \frac{
      \left[ \theta_{11}(z;\tau) \right]^2}{
      \left[ \eta(\tau) \right]^3
    } ,
  \end{gather}
  where
  \begin{equation}
    \mu(z;\tau)=
    \frac{\I \, \E^{\pi \I z}}{
      \theta_{11}(z;\tau)
    }
    \sum_{n \in \mathbb{Z}}
    (-1)^n \,
    \frac{q^{\frac{1}{2} n (n+1)} \, \E^{2 \pi \I n z}}{
      1 - q^n \, \E^{2 \pi \I z}
    }  .
  \end{equation}
  
\item non-BPS (massive) representations ($h>\frac{1}{4}$)
  \begin{equation}
    \ch^{\widetilde{R}}_{h > \frac{1}{4} ,\ell={1\over 2}}(z;\tau)
    =
    q^{h-{3\over 8}} \, {\left[ \theta_{11}(z;\tau) \right]^2\over
      \left[ \eta(\tau) \right]^3}.
  \end{equation}

\end{itemize}

\end{document}